\documentclass[aps,pra,twocolumn]{revtex4}

\usepackage{graphicx}
\usepackage{hyperref}
\usepackage{color}
\usepackage{amsmath,amssymb}
\usepackage[utf8]{inputenc}
\usepackage{dcolumn}

\begin{document}

\title{Trapping, cooling, and photodissociation analysis of state-selected H$_2^+$ ions produced by (3+1) multiphoton ionization}

\author{Julian Schmidt$^1$, Thomas Louvradoux$^1$, Johannes Heinrich$^1$, Nicolas Sillitoe$^1$, Malcolm Simpson$^2$, Jean-Philippe Karr$^{1,3}$, and Laurent Hilico$^{1,3}$}
\affiliation{$^1$Laboratoire Kastler Brossel, Sorbonne Université, CNRS, ENS-PSL Research University, Collège de France\\
4 place Jussieu, F-75005 Paris, France}
\affiliation{$^2$Institut für Ionenphysik und Angewandte Physik, Universität Innsbruck, Technikerstraße 25, 6020 Innsbruck, Austria}
\affiliation{$^3$Université d'Evry-Val d'Essonne, Université Paris-Saclay, Boulevard François Mitterrand, F-91000 Evry, France}
\date{\today}

\begin{abstract}
We report on the production of cold, state-selected H$_2^+$ molecular ions in a linear RF trap. The ions are produced by (3+1) resonance-enhanced multi-photon ionisation (REMPI) of H$_2$, and sympathetically cooled by laser-cooled Be$^+$ ions. After demonstrating and characterizing the REMPI process, we use photodissociation by a deep UV laser at 213~nm to verify the high vibrational purity of the produced H$_2^+$ ion samples. Moreover, the large difference between the photodissociation efficiencies of ions created in the $v=0$ and $v=1$ levels provides a way to detect a $v=0 \to 1$ transition. These results pave the way towards high-resolution vibrational spectroscopy of H$_2^+$ for fundamental metrology applications.
\end{abstract}

\maketitle

\section{Introduction}\label{sect-introduction}

Precision spectroscopy of hydrogen molecular ions (HMI) is a promising method to improve the determination of fundamental constants~\cite{Wing76}. By measuring several well-chosen rovibrational transitions in both H$_2^+$ and HD$^+$, it is possible to determine the $m_p/m_e$ and $m_d/m_e$ mass ratios, the proton and deuteron charge radii and the Rydberg constant~\cite{Karr16}. In particular, the present theoretical uncertainty of rovibrational transition frequencies in HMI, which is at a few parts-per-trillion level~\cite{Korobov17}, allows for significant progress in the determination of $m_p/m_e$~\cite{Heisse17,Sturm14}. H$_2^+$ also has strong potential for ultra high resolution spectroscopy at the 10$^{-17}$ level for tests of the time independence of $m_p/m_e$~\cite{Schiller14,Karr14}.

So far, dipole-allowed rovibrational transitions have been measured in sympathetically cooled HD$^+$ ion ensembles with 1-2~ppb accuracies, limited by Doppler broadening~\cite{Koelemeij07,Bressel12,Biesheuvel16}. More recently, a pure rotational transition was measured in the Lamb-Dicke regime with a $0.38$~ppb accuracy~\cite{Alighanbari18,Alighanbari20}, and a two-photon vibrational transition was measured with a 2.9~ppt uncertainty~\cite{Patra20}.

Spectroscopy of the homonuclear species H$_2^+$ poses additional challenges in comparison to HD$^+$, due to the fact that rovibrational transitions are dipole-forbidden. In the case of HD$^+$, the ions may be simply produced by electron-impact ionization, after which spontaneous decay to the ground vibrational state occurs within a few hundreds of milliseconds. The population of a rotational state of interest for spectroscopy can be further increased using rotational cooling methods~\cite{Bressel12,Schneider10} or exploiting rotational transitions induced by blackbody radiation (BBR)~\cite{Biesheuvel16,Biesheuvel17}. In H$_2^+$, due to the very long lifetimes of rovibrational levels in the $10^6$~s range~\cite{Posen83}, ions created by electron-impact remain scattered in many ro-vibrational states, which strongly limits the achievable spectroscopic signal~\cite{Karr12}. In addition, methods relying on optical pumping or BBR-induced transitions are not available.

One possible way of producing H$_2^+$ ion ensembles in a pure rovibrational state is to create the ions by electron impact, and achieve rovibrational cooling through collisions with a cold He buffer gas~\cite{Hansen14}, as recently analyzed theoretically by S.~Schiller and coworkers~\cite{Schiller17,Hernandez17a,Hernandez17b}. They estimated that cooling to the vibrational ground state, in competition with loss of molecules through the reaction H$_2^+$+He $\rightarrow$ HeH$^+$ + H, should occur within a time of about 10~s, and that rotational cooling within the $v=0$ state is feasible within approximately 40~s for a He density of $10^{9}$~cm$^{-3}$. Since collisions with He do not affect the nuclear spin, in order to get a single rotational state one should in addition start from para-H$_2$ gas using an ortho-para H$_2$ converter, or remove ions in the unwanted $L=0$ (para) or $L=1$ (ortho) state by resonance-enhanced multiphoton dissociation (REMPD).

In this paper, we demonstrate an alternative method relying on state-selective production of H$_2^+$ by four-photon (3+1) resonance-enhanced multiphoton ionization (REMPI) of H$_2$~\cite{OHalloran87}. We report on the production, trapping, and sympathetic cooling of state-selected H$_2^+$ ions in $v=0$ and $v=1$. A similar idea has been used to produce cold, state-selected N$_2^+$ ions by threshold ionization~\cite{Tong10}. This work is an important step towards our goal of performing high-resolution spectroscopy of the $(v=0,L=2) \to (v'=1,L'=2)$ Doppler-free two-photon transition~\cite{Karr12}. The transition will be detected by REMPD, similarly to previous experiments in HD$^+$, using a 213~nm laser to selectively dissociate ions excited to the $v=1$ level. Here, we use photodissociation to verify the vibrational selectivity of the REMPI process. Moreover, we show that the $v=1$ level is dissociated much more efficiently than $v=0$, confirming the possibility of detecting a $v=0 \to v'=1$ transition via REMPD.

The paper is organized as follows: after describing the experimental setup in Sec.~\ref{sect:setup}, we characterize REMPI H$_2^+$ ion production in the vibrational states $v=0$ and $v=1$ in Sec.~\ref{sect:REMPI}. The sympathetic cooling of the state-selected H$_2^+$ ions, is addressed in Sec.~\ref{sect:sympacool}, where we also describe a nondestructive method to measure the ion number. Finally, experimental results of state-selective photodissociation of H$_2^+$ are presented and analyzed in Sec.~\ref{sect:photodiss}.

\section{Experimental setup} \label{sect:setup}

\begin{figure*}
\centering
\includegraphics[width=0.96\textwidth]{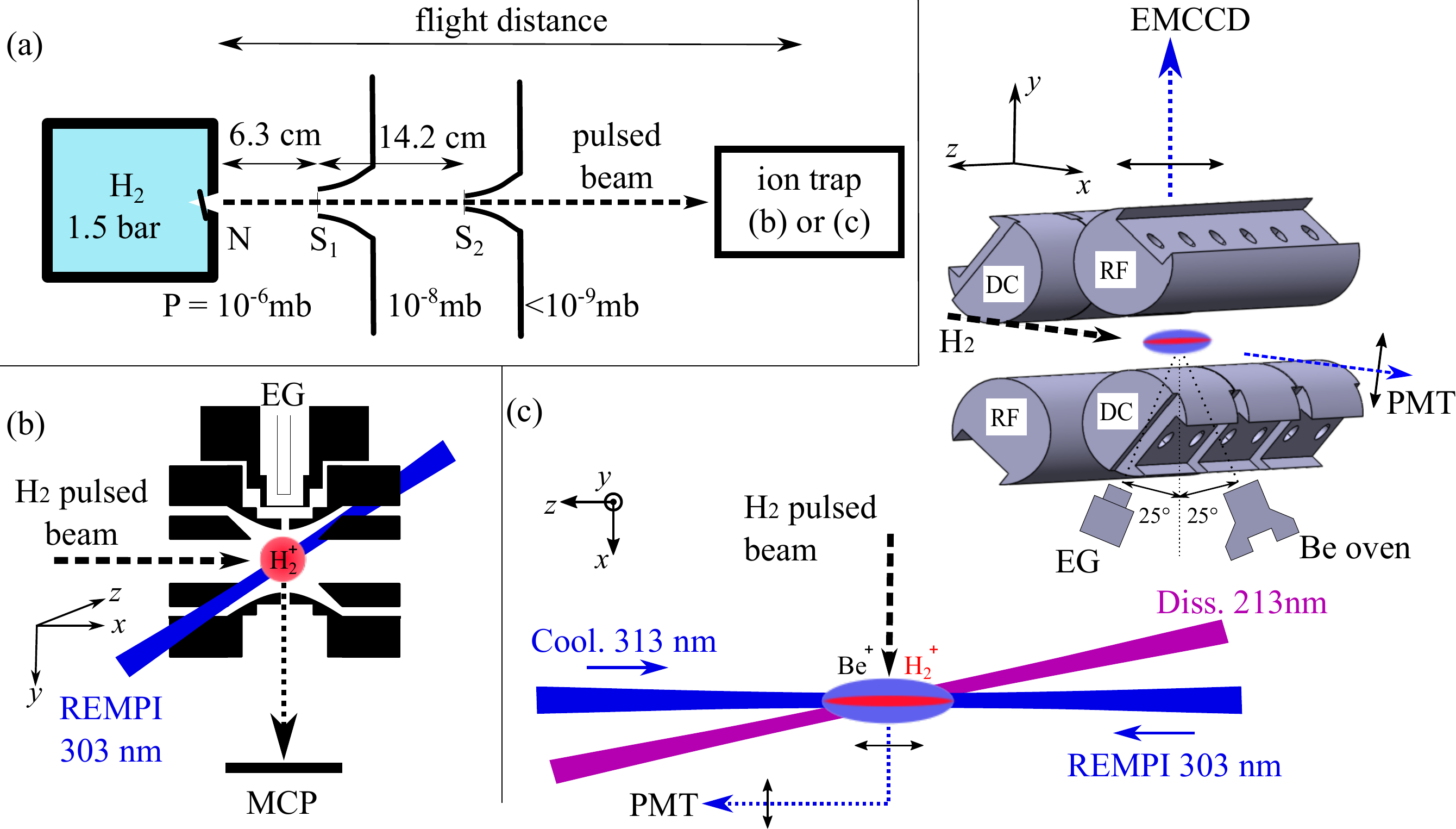}
\caption{Schematic of the experimental setups. (a) The molecular beam setup, which can be connected to one of two UHV chambers containing different ion traps. N: pulsed valve with a 150~$\mu$m nozzle, S$_1$: 1~mm skimmer, S$_2$: 150~$\mu$m skimmer. Typical pressures in the different pumping stages are indicated. The pressures in the ion traps are typically $10^{-10}$~mbar. (b) Hyperbolic RF ion trap. EG: electron gun (home-made tungsten hair pin), MCP: microchannel plate. (c) Segmented linear RF ion trap with laser-cooled Be$^+$ ions. EG: Kimball Physics electron gun FRA-2X1-2. Imaging of the Be$^+$ fluorescence on the EMCCD camera is done using a $f=70$~mm objective with a magnification of 9. Another part of the fluorescence is collected by an in-vacuum lens ($f=30$~mm) and focused on a photomultiplier tube (PMT).}\label{fig:sketch}
\end{figure*}

The experimental setup consists of a pulsed H$_2$ molecular beam apparatus (Fig.~\ref{fig:sketch}(a)) which can be connected to two different ultrahigh vacuum (UHV) chambers, one containing a hyperbolic RF ion trap (Fig.~\ref{fig:sketch}(b)) and the other one a linear RF ion trap (Fig.~\ref{fig:sketch}(c)). A pulsed, tunable laser with a wavelength near 300~nm (called ''REMPI laser'' in the following) can be used to create H$_2^+$ ions by REMPI in either trap.

\subsection{H$_2$ molecular beam}
The molecular beam is produced by a fast pulsed piezo valve (Amsterdam Piezo Valve) with a 150~$\mu$m nozzle and a 5-10~$\mu$s opening time from a reservoir with a stagnation pressure of 1.5~bar. The beam passes through a first skimmer with aperture 1~mm, located 6.3~cm from the nozzle, and a second skimmer with aperture 150~$\mu$m, located 20.5~cm from the nozzle. The pulsed valve chamber is evacuated with a CF160 high compression ratio turbo pump. The chamber between the two skimmers is evacuated using a CF63 turbo pump and an ion-getter pump in order to maintain ultrahigh vacuum conditions in the ion trap chamber. An all-metal gate valve allows for separating the molecular beam chamber and mounting it to either the hyperbolic ion trap chamber or to the linear ion trap chamber. The two traps are described in more detail below. The distance between the nozzle and the trap center (''flight distance'' in Fig.~\ref{fig:sketch}(a)) is 62(1)~cm for the hyperbolic trap and 65(1)~cm for the linear trap. Finally, the orientation of the molecular beam with respect to the ion trap can be adjusted by means of a universal joint (not shown in Fig. 1) placed between the second skimmer (S$_2$) and the gate valve.
\subsection{Hyperbolic trap}
The hyperbolic trap~\cite{Karr12} is made of stainless steel with a ring of inner radius $r_0=4.2$~mm and two endcaps separated by $2z_0=6$~mm. It is operated with an RF voltage amplitude of 180~V at 13.4~MHz and DC voltage of 1~V between the ring and endcap electrodes. The upper and lower endcap electrodes feature along the $y$ axis, respectively, a 1~mm hole for electron injection from a home-made electron gun polarized at -150~V, and a 3~mm hole for ion extraction and detection on a microchannel plate. The H$_2$ molecular beam and the REMPI laser beam are perpendicular to each other, and respectively aligned with the $x$ and $z$ direction. They enter the trap along two pairs of 5~mm holes drilled in the ring electrode.
\subsection{Linear trap}
The segmented linear trap is made of four cylindrical molybdenum rods of 4~mm radius and 36~mm length, and the distance from the trap axis to the electrodes is $r_0=3.5$~mm. Two diagonally opposing rods are segmented into 12~mm long parts. The trap is operated with an RF amplitude of up to 550~V at 13.3~MHz and a DC voltage up to 8~V between the central part and ends of the segmented rods, allowing for trapping of both H$_2^+$ and Be$^+$. 
The Be$^+$ ions are created via electron-impact ionization of a neutral Be effusive beam coming from a ceramic Be oven described in~\cite{Heinrich2018}. A Be wire (length 5~mm, diameter 0.25~mm) is inserted in a four hole ceramic tube (tube diameter 1.2 mm, hole diameter 0.254 mm)  heated by a 0.25~mm diameter tantalum wire wound through the three other holes. The Be oven is located at 46 mm from the trap center. The Be effusive beam is collimated by a 1 mm diaphragm located at 25~mm from the oven resulting in a 20~mrad beam half-angle and a beam diameter of about 2~mm at the trap center. The oven current is 3~A, slightly above the threshold current (2.9~A) to observe Be$^+$ ions in the trap.
An electron gun (Kimball Physics FRA-2X1-2) is installed with a 20~mm working distance from the trap center. The electron gun filament is polarized at -110~V with respect to ground potential, and the filament current is set to 1.4~A to have a 10~$\mu$A emission current. Note that due to the RF electric field in the trap, the electron beam energy and spot size cannot be known precisely.
The oven and electron gun are oriented symetrically with respect to the vertical $y$ axis, making a 25 degree angle with it and aiming at the trap center (see Fig.~\ref{fig:sketch}(c)).
To create Be$^+$ ions in the trap, both the oven and electron gun are turned on for 30 s and then turned off. The typical trapping voltages are 470 V RF amplitude, 0~V on the central segment and 8~V on the endcap segments resulting in a stability parameter $q_x=0.12$ and the secular frequencies $f_x=f_y=550$~kHz in the $x$ and $y$ directions and $f_z=$126~kHz in the $z$ direction for Be$^+$. 

The Be$^+$ cooling laser  at 313~nm is aligned with the trap axis $z$. The REMPI laser is counterpropagating with respect to the cooling laser and is aligned using diaphragms centered on the cooling laser beam. The dissociation laser (described and used in Sec.~\ref{sect:photodiss}) propagates in the $x-z$ plane with an angle of $28.5^{\circ}$ with respect to the trap axis. The H$_2$ molecular beam enters the trap along the $x$ axis. The fluorescence of the Be$^+$ ions is imaged on an electron multiplying charge-coupled device (EMCCD) camera along the vertical $y$ axis and focused on a photomultiplier tube (PMT) along the horizontal $x$ axis. The background pressure in both trap chambers is of the order of $10^{-10}$~mbar.
\subsection{REMPI laser}
The REMPI laser is a frequency-doubled pulsed dye Laser (Sirah Precision scan) pumped with a Q-switched YAG laser with a repetition rate of 20~Hz and a pulse energy of 110~mJ. We get about 4~mJ at 303~nm using an optimized mix of rhodamine 610 (64\%) and rhodamine 640 (36\%) with 0.12~g/l total concentration in ethanol, and 3~mJ at 296~nm using pure rhodamine 610. The elliptical beam profile is circularized using cylindrical lenses. It has a divergence of 1.6~mrad and an M$^2$ factor of about 2.

\section{Creation of H$_2^+$ ions by REMPI}\label{sect:REMPI}

Creation of hydrogen molecular ions from the pulsed H$_2$ molecular beam (Fig.~\ref{fig:sketch}(a)) by REMPI was first tested and optimized in the absence of sympathetic cooling, using the hyperbolic rf trap (Fig.~\ref{fig:sketch}(b)). With this trap, production yields can be easily measured by extracting the ions towards a multichannel plate detector. In addition, the experimental cycle is shorter (about 7~s) than in the linear trap (about 45~s) since it does not include the preparation of a cold Be$^+$ crystal. This makes the hyperbolic trap well suited to measure the REMPI spectrum and optimize the molecular beam and laser parameters.

(3+1) REMPI of H$_2$ is a two-step process consisting of a resonant three-photon excitation to a given rovibrational state of the Rydberg $C ^1\Pi_u$ electronic state, followed by nonresonant one-photon ionization. It thus requires tight focusing of the REMPI laser.
The laser beam is expended to a diameter of about 12~mm and focused with a $f=189$~mm spherical lens. A spot diameter of about 12~$\mu$m is expected, but the Rayleigh length  $Z_R=0.2$~mm is small as compared to the molecular beam diameter at the trap center or to the precision of the focal spot localization along the beam axis. The effective beam diameter in the interaction region is thus larger than 12~$\mu$m.
The very small ion production volume leads to low ionization yields, and precise overlap of the REMPI beam with both the trap center and the molecular beam is mandatory in order to observe a signal. Overlap with the trap center is achieved using the photodissociation of H$_2^+$ ions created by electron-impact ionization. This process populates a large number of vibrational levels (from $v=0$ to 12)~\cite{Busch72}, and a significant fraction of the produced ions is efficiently dissociated at 303~nm. The beam is then aligned with the trap center by maximizing the fraction of photodissociated ions, after which the focal point is located at the center by {\em minimizing} the photodissociated fraction. Finally, the electron gun is turned off and the molecular beam is aligned with the trap center using the universal joint mentioned in Sec.~\ref{sect:setup}, in order to optimize the REMPI signal.

\begin{figure}[h!]
\centering
\includegraphics[width=0.48\textwidth]{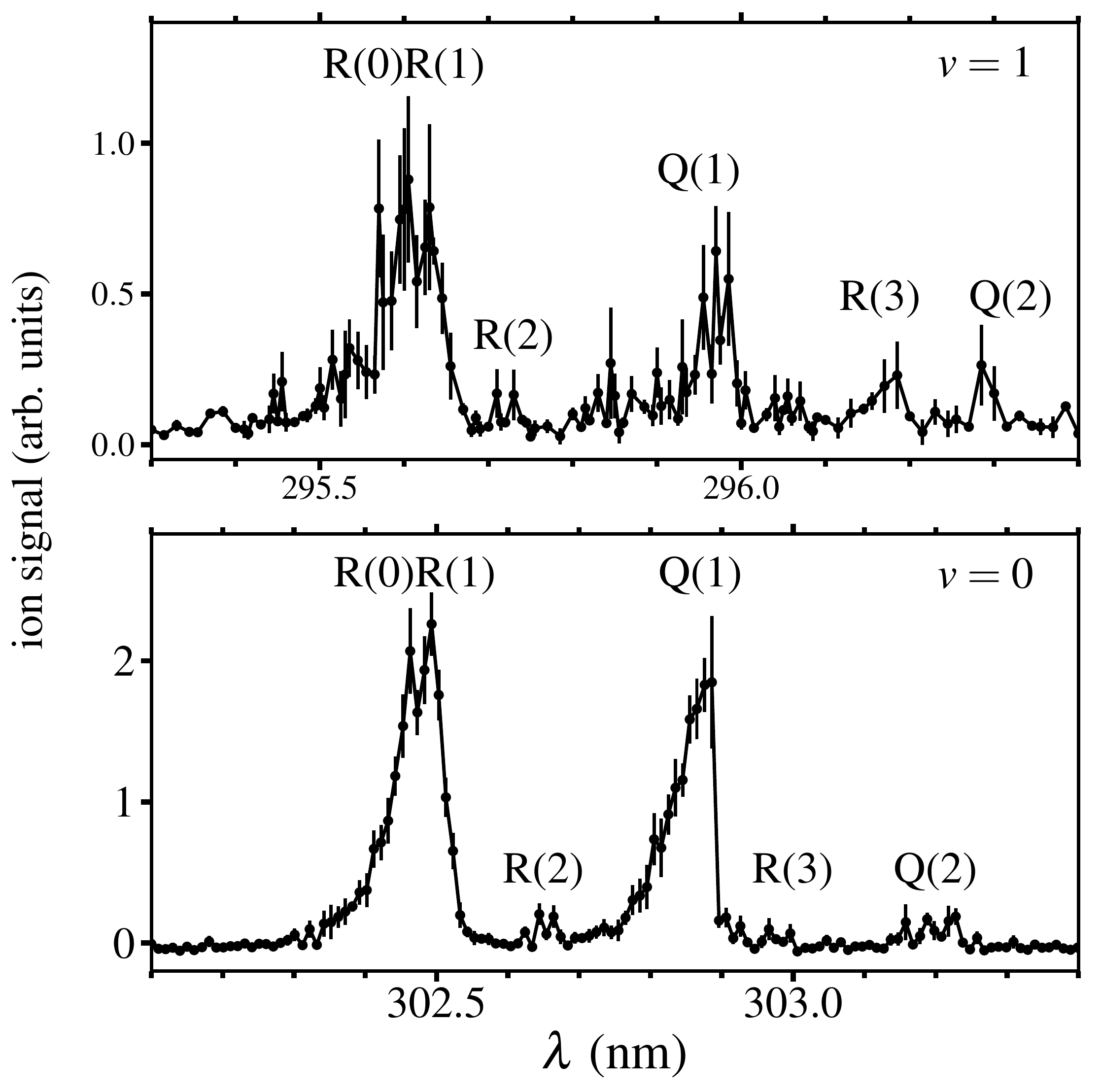}
\caption{Mean ion signal as a function of the REMPI laser wavelength, obtained after 3~s of REMPI pulses synchronized with H$_2$ pulses (corresponding to 60 pulses). The ion signal is averaged over 10 to 20 experimental cycles. Error bars represent the standard deviation. The positions of the resonances are in good agreement with the data shown in~\cite{Pratt84}, allowing their identification.}\label{fig:rempi_spectrum_both}
\end{figure}

Spectra of the REMPI process near 303 nm and near 296 nm, are shown in Fig.~\ref{fig:rempi_spectrum_both} and are in good agreement with previous results~\cite{Pratt84}. 
The signal in the $v=1$ state is weaker due to several reasons: reduced laser power, different transition strengths and possible dissociation of the ions it produces. The $v=1$ photodissociation cross section at 296~nm is 2.18$\times 10^{-26}$m$^2$~\cite{Busch72} leading to the dissociation of up to 80\% of the produced ions for 3 mJ, 7 ns pulses focused with a 12~$\mu$m spot diameter. The $v=1$ signal is only suppressed by a factor of two, which indicates that the spot diameter of 12~$\mu$m is underestimated and that the effective diameter is at least two times larger. 
We observe an asymmetry and a broadening of the lines with an FWHM of about 0.1~nm. We attribute them to the AC Stark shift of the intermediate $C$ state of H$_2$~\cite{Posthumus08,Allendorf91}. An upper bound of this shift is given by the ponderomotive shift of the loosely bound electron at the center of the beam,
\begin{equation}
 \frac{q^2\lambda^2 E}{4\pi^3\epsilon_0mc^3w_0^2\tau}=8.7\,\text{meV} ,
\end{equation}
where $q$ is the elementary charge, $m$ the electron mass, $w_0=6\ \mu$m and where $E=4$~mJ and $\tau=7$~ns are the typical pulse energy and duration. This corresponds to a 0.64~nm shift in terms of wavelength, which is significantly larger than the observed maximum shift, indicating again that the effective beam diameter is at least two times larger than expected.

For the experiments described in the following, we tune the wavelength to $302.47$ nm or $295.61$ nm to produce ions in the $v=0$ or the $v=1$ state via the combination of the (unresolved) R(0) and R(1) lines.

\begin{figure}[h!]
\includegraphics[width=0.48\textwidth]{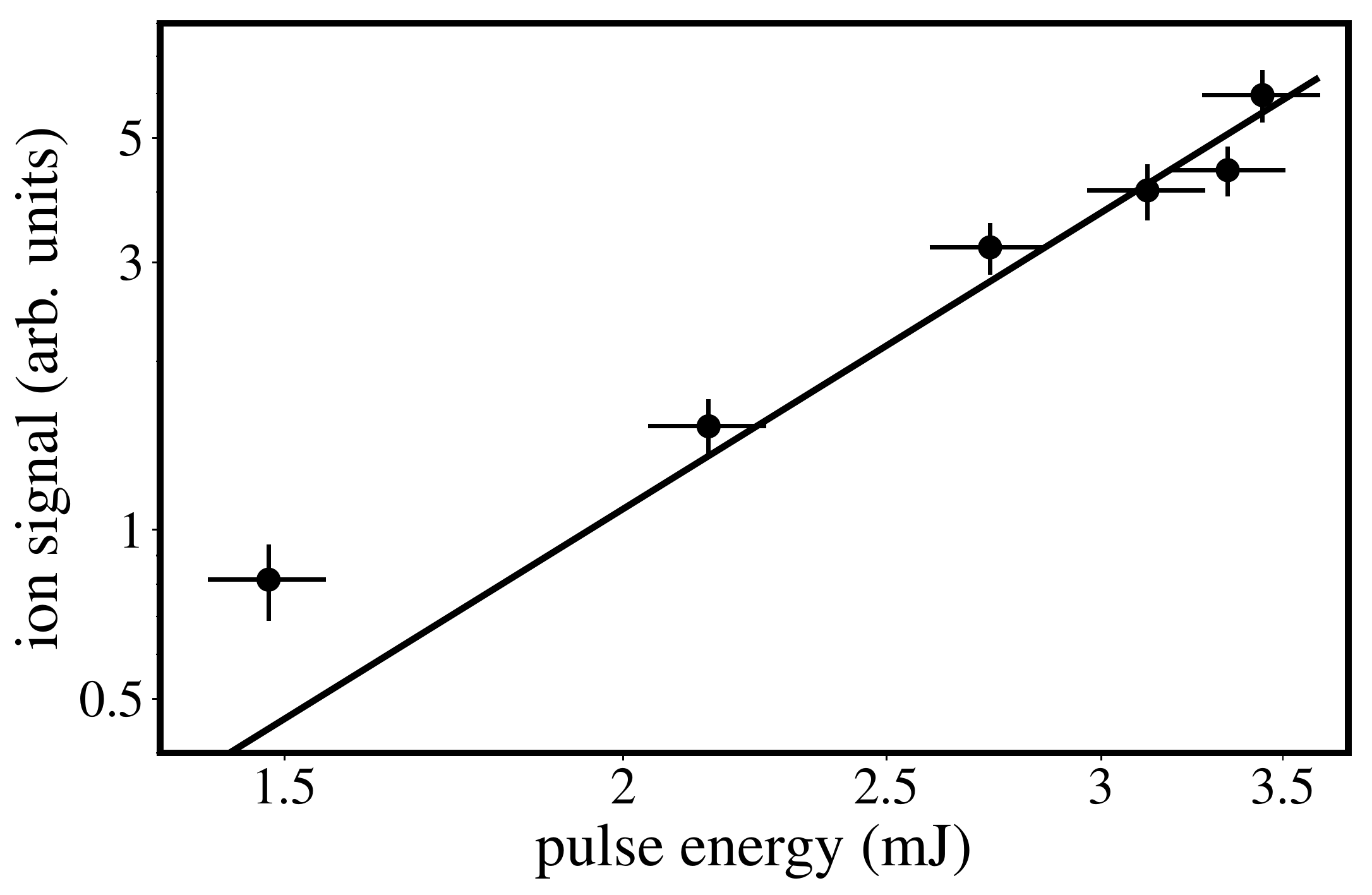}
\caption{Mean ion signal from the $v = 0$, R(0)-R(1) line (see Fig. 2) as a function of pulse energy. Each data point is an average of 10 to 60 experiments. Horizontal error bars represent a 5\% uncertainty of the pulse energy. Vertical error bars include the statistical deviation within one measurement run and a conservative estimate of the stability over the duration of the entire experiment. The stability is in part limited by the fluctuating $H_2$ background gas pressure. The solid line is a fit with a cubic model, ion signal $= 0.136 \, E^3$.}\label{fig:cubic_dependence}
\end{figure}

Fig.~\ref{fig:cubic_dependence} shows the dependence of the REMPI signal on the pulse energy. The data is in good agreement with a cubic law, indicating that only the ionization step from the $C$ state is saturated but not the three-photon transition rate. Saturation of the REMPI process was observed by another group using a pulse energy of approximately 20~mJ~\cite{Urbain}.

\begin{figure}[h!]
\includegraphics[width=0.48\textwidth]{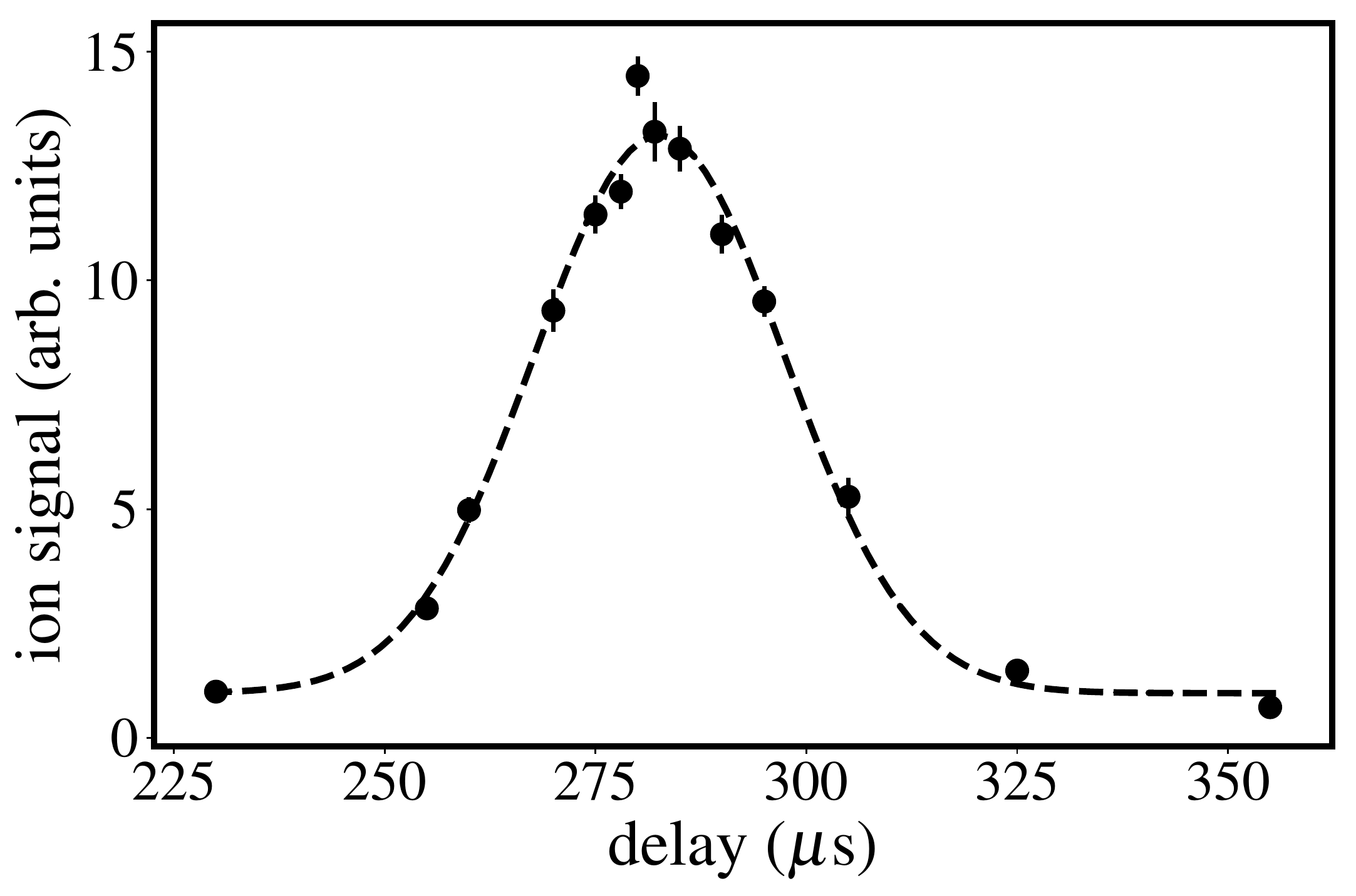}
\caption{Mean ion signal in the $v=0$ state as a function of the time of flight, i.e. the delay between the opening of the pulsed valve and the REMPI laser pulse. Each data point is the average of 10 to 20 experimental sequences. Error bars represent the standard deviation. The backing pressure is 2 bar and the valve opening duration is estimated to be shorter than 10 $\mu$s. The dashed line is a fit by a Gaussian function with a FWHM of 35(2)~$\mu$s.}\label{fig:delay}
\end{figure}

Fig.~\ref{fig:delay} shows the dependence of the REMPI signal on the delay between the opening of the pulsed valve and the laser pulses. The optimum observed for $\Delta t_{\text{tof}} = 282(1) \, \mu$s, for the 62(1)~cm distance between the nozzle and the trap (see Fig.~\ref{fig:sketch}), corresponds to a most probable velocity of $v_{mp}=2190(40)$ m/s. The initial pulse duration is difficult to access experimentally. By observing the drive voltage of the pulsed valve, we estimate it to be shorter than 10~$\mu$s. Neglecting this duration, the measured FWHM of 35(2)~$\mu$s after expansion corresponds to a velocity dispersion smaller than $\delta v=271(15)$ m/s. These values indicate that the molecular beam is operated in an intermediate regime between an effusive beam, for which $v_{mp}=1906$~m/s and $\delta v=1780$~m/s, and a supersonic beam, for which $v_{mp}=\sqrt{7 RT_0/M}=2900$~m/s (for a diatomic molecule) with a narrow velocity distribution~\cite{Irimia09} . In the latter formula, $R$ is the molar gas constant, $M$ the molar mass and $T_0$ the room temperature.

During injection of the pulsed molecular beam, the pressure in the ion trap chamber remains at an acceptably low value. After a 3~s REMPI cycle, corresponding to 60 gas pulses, it increases by about 1.~10$^{-10}$~mbar for a few seconds. Over tens of experimental cycles, the pressure eventually increases to a few 10$^{-10}$~mbar. We believe this stems from a slow increase of the pressure in the intermediate vacuum sections of the molecular beam apparatus (see Fig.\ref{fig:sketch}) rather than from the gas flux of the H$_2$ pulses themselves. Faster pumping of the intermediate vacuum may alleviate this issue. We also note that using pure para-H$_2$ would allow the amount of injected gas to be reduced by a factor of three.

\section{Sympathetic cooling of state-selected ions and ion counting}\label{sect:sympacool}

After testing state-selective production with the hyperbolic trap, we now use the linear rf trap where the H$_2^+$ ions produced by REMPI can be sympathetically cooled by laser-cooled Be$^+$ ions \cite{Molhave00,Willitsch12}.

The experimental sequence begins by creating a Coulomb crystal of several hundred Be$^+$. The ions are produced as explained in section~\ref{sect:setup} and laser cooled at a detuning of about $5\Gamma=2\pi\times$~100~MHz and an intensity of about $1 \times I_{\text{sat}}$=0.83~mW/mm$^2$. The cooling laser at 313~nm is obtained by sum frequency mixing of two fiber lasers at 1051 and 1549~nm amplified up to 4.5~W and cavity enhanced second harmonic generation in a Brewster-cut 12~mm long BBO non-linear crystal~\cite{Wilson11}. 

Ionic species heavier than Be$^+$ are also created by the electron impact process from atoms or molecules evaporated from the Be oven. These are washed out of the trap by adding for 1~s a DC component of about 9V to the RF voltage, which makes singly charged ions of mass $\geq$10 unstable. The RF voltage amplitude is then reduced to 340~V to lower the $q$ stability parameter to 0.09  and the secular frequency to $f_x=400$~kHz for Be$^+$, in order to keep H$_2^+$ stability parameter below 0.4.

\begin{figure}[h!]
\centering
\includegraphics[width=0.48\textwidth]{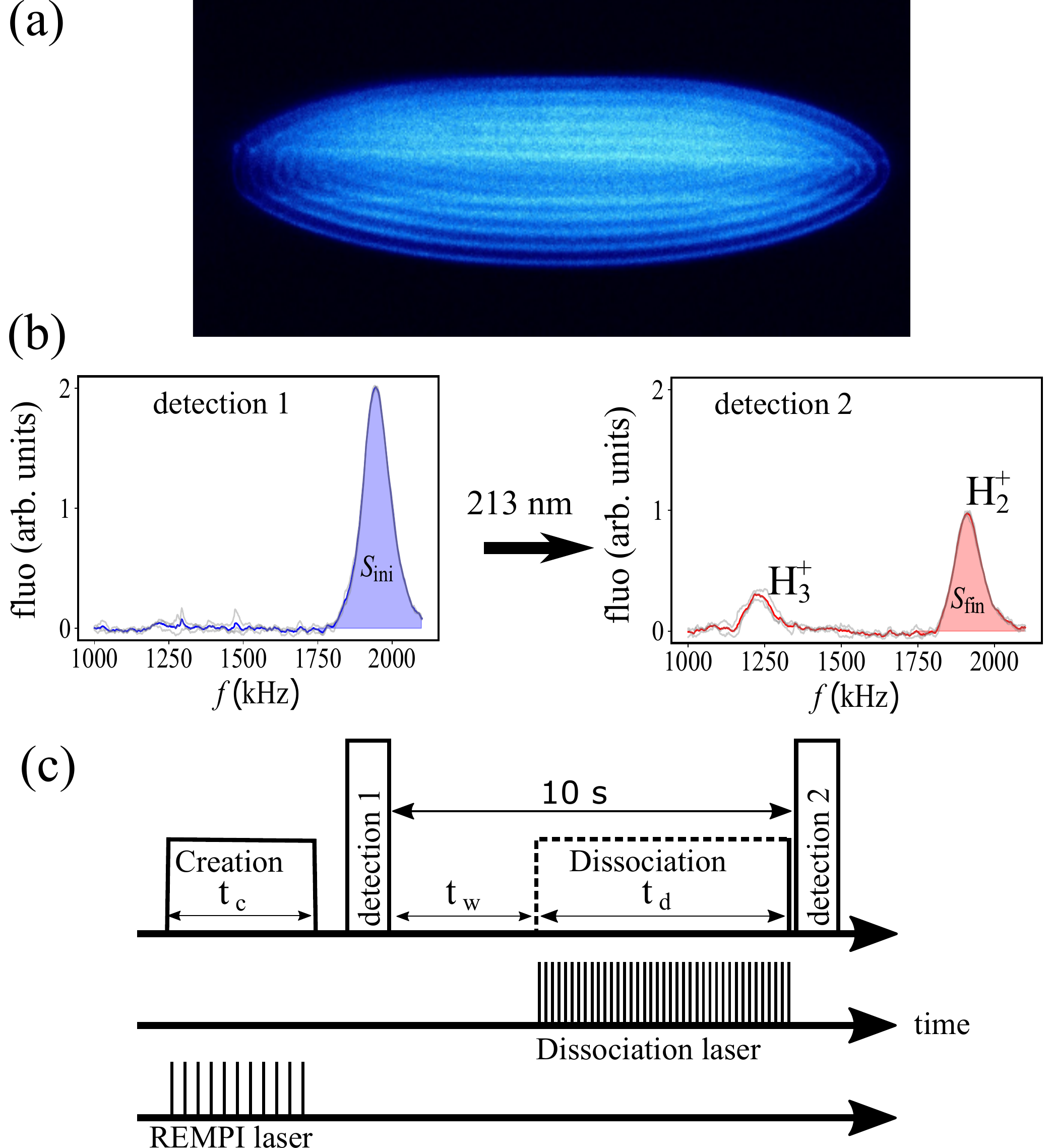}
\caption{
Sympathetic cooling and detection of H$_2^+$ ions produced by REMPI. (a) Fluorescence image of a Be$^{+}$ ion Coulomb crystal in the linear trap. The dark region along the horizontal axis is due to the presence of sympathetically cooled H$_2^+$ ions. (b) Typical detection signals of H$_2^+$ ions. The fluorescence of the Be$^{+}$ ions is measured as a function of the excitation frequency $f$. Three 0.5~s sweeps (light grey) are averaged. The peak at about 1.8~MHz corresponds to excitation of the radial secular motion of H$_2^+$. In the second signal, the peak around 1.2~MHz corresponds to the appearance of H$_3^+$ ions due to chemical reactions of H$_2^+$ with H$_2$ (see text). The H$_2^+$ ion signal is the shaded area below the curve. (c) Experimental sequence for measuring the photodissociation efficiency. The H$_2^+$ ion number is measured non-destructively before and after interaction with the 213~nm laser.
}
\label{fig:sympacool}
\end{figure}

Next, the pulsed molecular beam, synchronized with the REMPI laser, is turned on for a variable time between $0.5\,\text{s}<t_c<3\,\text{s}$ to create state-selected H$_2^+$ ions. The focusing of the REMPI laser slightly differs from Sec.~\ref{sect:REMPI}: the beam is expanded to about 16~mm and focused using a $f=236$~mm lens, which again results in an expected spot diameter of about 12~$\mu$m. After the pulsed laser is turned off, we observe dark ions embedded in the Coulomb crystal, see Fig.~\ref{fig:sympacool}(a).

The dark ion species can be identified by exciting their secular motion in the trap and observing the corresponding effect on the fluorescence of the Be$^+$ ions. This signal also provides a way to measure nondestructively the number of trapped H$_2^+$ ions, which will be used for the REMPD spectroscopy of H$_2^+$, as done in previous experiments on HD$^+$, see e.g.~\cite{Koelemeij07,Biesheuvel16,Biesheuvel17}. In practice, an AC voltage is applied to one of the trap electrodes and its frequency is swept linearly with time. When only Be$^+$ is trapped, we observe an increase in the fluorescence rate near $f_{\text{sec,Be}^+} = 400$~kHz. After REMPI and when dark ions are present, we observe a peak near $4.5\times f_{\text{sec,Be}^+} = 1800$~kHz, indicating the presence of H$_2^+$, see Fig.~\ref{fig:sympacool}(b). An additional peak appears near 1200~kHz after 10-20~s, which we attribute to the formation of H$_3^+$ through the reaction H$_2^+$+H$_2$ $\rightarrow$ H$_3^+$+H. The observed H$_3^+$ formation rate is proportional to the H$_2^+$ ion number and is consistent with a partial H$_2$ pressure of about $5\times 10^{-10}$~mbar~\cite{Meir19}. The area under the peak can be assumed, as a reasonable approximation, to be proportional to the number of H$_2^+$ ions, as analyzed in detail in~\cite{Biesheuvel17}.

\section{Photodissociation of state-selected, sympathetically cooled H$_2^+$ ions} \label{sect:photodiss}

The study of photodissociation of the H$_2^+$ ions created by REMPI serves two purposes. Firstly, due to the strong dependence of the photodissociation cross-section on the vibrational state, the photodissociation dynamics gives information on the vibrational population distribution, and thus allows us to check the vibrational selectivity of the REMPI process. Secondly, we will show evidence of the difference in the dissociation efficiencies of the $v=0$ and $v=1$ vibrational states, which demonstrates the possibility of detecting a $v=0 \to v'=1$ transition by REMPD.

We use a pulsed laser at $\lambda_{d} = 213$~nm (Xiton Photonics Impress 213) with average power up to $150$~mW, repetition rate $f_\text{rep} = 12.5$~kHz and a beam waist radius of 300~$\mu$m at the trap center (see Fig.~\ref{fig:sketch}(a)). The experimental sequence is depicted in Fig~\ref{fig:sympacool}(c). After creation and sympathetic cooling of H$_2^+$ as described in the previous Section, we apply the photodissociation laser for the variable duration $\Delta t_{d} = 0...10$~s. To ensure that losses related to background gas collisions are identical for data points with different $\Delta t_{d}$, we add a waiting time $\Delta t_{w} = 0...10$~s before $\Delta t_{d}$, such that the total duration of one experiment is always $\Delta t_{w} + \Delta t_{d} = 10 \, \text{s}$. This does not account for possible losses due to the pressure increase stemming from the laser pulses themselves. However, in contradistinction with previous results~\cite{Karr12}, such losses are negligible here because we use much lower UV powers ($P_d \sim 10-20$~mW). This is partly due to the deeper UV wavelength (213~nm instead of 248~nm in Ref.~\cite{Karr12}), for which the photodissociation cross-section of the $v=1$ level is much higher. Another reason is that the sympathetically cooled H$_2^+$ ions are localized in a smaller volume in the vicinity of the trap center, allowing a better overlap with the photodissociation laser.

We measure the ion signal before ($S_{\text{ini}}$) and after ($S_{\text{fin}}$) applying the photodissociation laser, as explained in Sec.~\ref{sect:sympacool}. From this we deduce the survival probability $p_{s}(\Delta t_d,\eta) = S_{\text{fin}}/S_{\text{ini}}$ after a given photodissociation duration $\Delta t_d$. In order to remove the effect of ion losses, the experiment is repeated without dissociation laser to measure the background survival probability $p_{\text{bg}}$, and we finally obtain a normalized signal $F(\Delta t_d,\eta)=p_{s}(\Delta t_d,\eta)/p_{\text{bg}}$.

These measurements have to be repeated to get sufficient statistics. In order to reduce the measurement time, Be$^+$ ions are loaded only when necessary. We proceed in the following way: at the end of the experimental sequence, any remaining H$_2^+$ and H$_3^+$ ions are removed from the trap by resonant excitation of the secular motion. This way, we obtain a pure Be$^+$ crystal and can repeat the experiment by again creating H$_2^+$ via REMPI. When the number of Be$^+$ ions has dropped below about half of its initial value, additional Be$^+$ ions are loaded into the trap via electron-impact ionization.

During operation of the molecular beam source, the Be$^+$ ions typically have a lifetime of 15 minutes. In addition, a large part of the losses occur through reactive formation of BeH$^+$ ions and can be reversed using photodissociation by the 213~nm laser~\cite{Sawyer15}.

\subsection{Model of the photodissociation dynamics}

Since the crystallized H$_2^+$ ions have very small motional amplitudes, they experience a nearly constant intensity during their interaction with the photodissociation laser. This allows a significantly simplified model of the photodissociation to be used compared to our previous work~\cite{Karr12}. For ions in the vibrational state $v$ and located at the beam focus, the survival probability after $N_p = \Delta t_d \times f_\text{rep}$ laser pulses is
\begin{equation}
 p_{d,v}(\Delta t_d,\eta) = \exp\left[-\eta N_p\frac{2 \lambda_d E_{p}\sigma_v }{\pi h c w_0^2}\right]\, .
\label{eq:remaining_frac}
\end{equation}
$E_{p}$ is the pulse energy (measured at the output of the laser), $\sigma_v$ the dissociation cross section for state $v$, $w_0 = 300$~$\mu$m the $1/e^2$ beam waist radius and $h$ is the Planck constant. The values of $\sigma_v$ for $v=0,1,2$ are $\{5.355,\,  386.3,\,  4787\}\times 10^{-25}$~m$^2$ at 303~nm~\cite{Busch72}. The parameter $\eta$ accounts for losses at 213~nm on the two mirrors used to steer the beam ($\approx$5\% each), losses on the uncoated focusing lens (8\%) and for imperfect overlap of the laser beam with the H$_2^+$.

To illustrate the latter point, let us consider an H$_2^+$ ion string of length $2 l = 1$~mm parallel to the trap axis. For perfect alignment, i.e. if the photodissociation beam axis crosses the middle of this string, in view of the 28.5$^{\circ}$ angle between the beam and the trap axis (see Fig.~\ref{sect:setup}) the overlap factor is expected to vary between $1$ at the center, and $e^{-2(l \sin(28.5^{\circ}))^2/w_0^2} = 0.28$ at each end of the string, with an average value approximately given by \[ \frac{1}{2l} \, \int_{-l}^{l} e^{-2(r \sin(28.5^{\circ}))^2/w_0^2} dr = 0.70. \] Including the effect of optical losses, one gets $\eta \simeq 0.92 \times 0.95^2 \times 0.70 = 0.58$. Note that the overlap factor may evolve during the depletion of the initial ion string via photodissociation. In the simple model presented here, we will assume $\eta$ to be fixed.

If the ions are in a mix of vibrational states with populations $n_v$, the overall survival probability is
\begin{equation}
p_{d}(\Delta t_d,\eta) = \sum_{v} n_v p_{d,v}(\Delta t_d,\eta) \, . \label{eq:diss_model}
\end{equation}

Due to collisions with background gas, the survival probability without the dissociation laser after $\Delta t_{w}=10$~s is $p_{\text{bg}}$, measured to be $\approx 85\%$. The survival probability with the dissociation laser is
\begin{equation}
p_{s}(\Delta t_d,\eta) = p_{\text{bg}} \times p_{d}(\Delta t_d,\eta),
\end{equation}
so that the normalized ion signal is
\begin{equation}
F(\Delta t_d,\eta) = p_{s}(\Delta t_d,\eta)/p_{\text{bg}} = p_{d}(\Delta t_d,\eta).
\end{equation}

\subsection{Experimental results}

\begin{figure}[h!]
\centering
\includegraphics[width=0.48\textwidth]{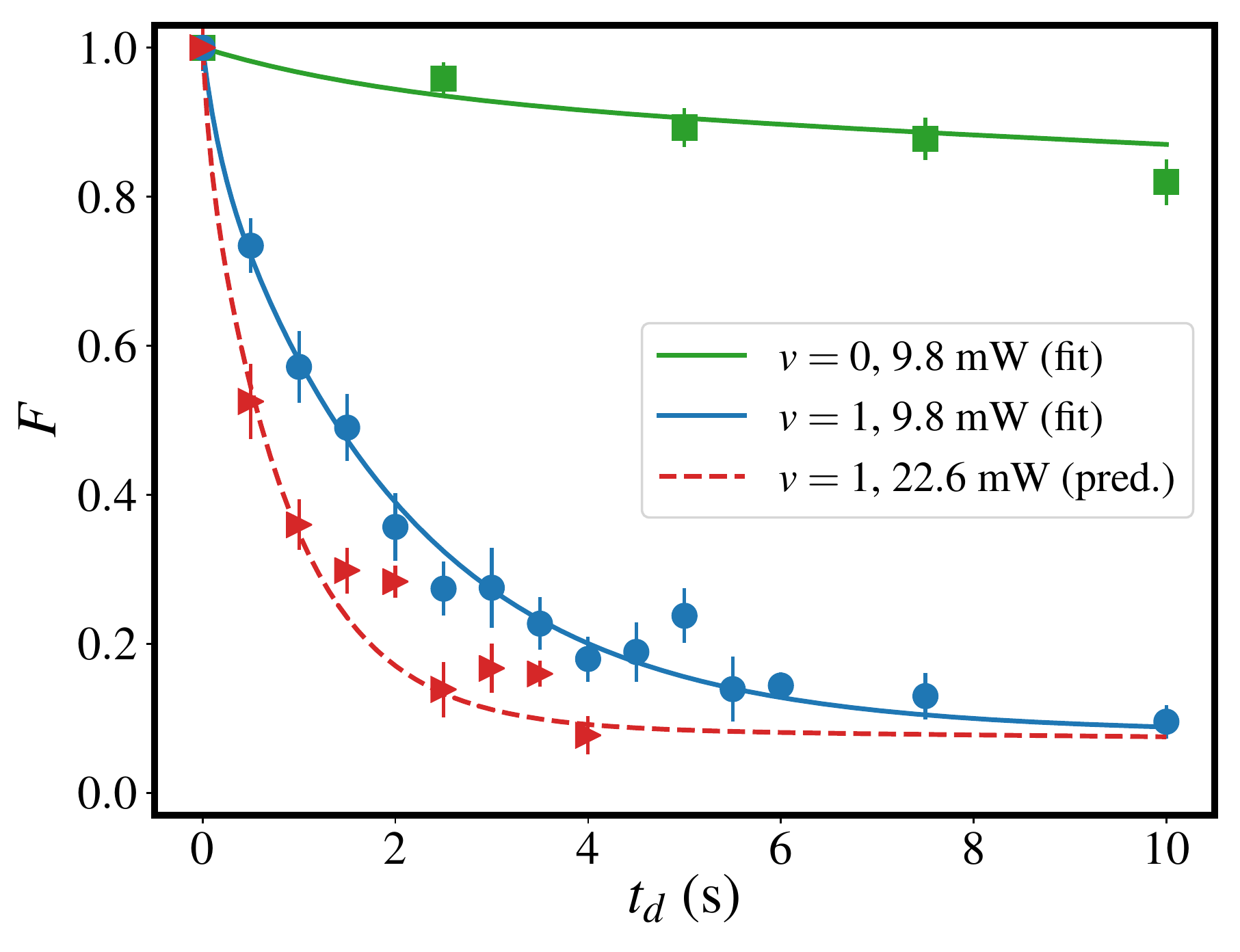}
\caption{
Dissociation of $H_2^+$ ions produced by REMPI in the $v=0$ and in the $v=1$ states and sympathetically cooled in the linear trap. The normalized remaining ion fraction $F$ (see text) is plotted as a function of the dissociation duration $t_d$ during which the 213~nm laser is applied. Each data point is an average of 6 to 25 measurements. Error bars correspond to the standard deviation. The solid lines indicate a common fit of the data with $P_d=9.8$~mW using Eq.~\ref{eq:diss_model}. The dashed line is a prediction for the data with $P_d=22.6$~mW. Squares: $\lambda = 302.47$~nm, $P_{d}=9.8$~mW, circles: $\lambda = 295.61$~nm, $P_{d}=9.8$~mW, triangles: $\lambda = 295.61$~nm, $P_{d}=22.6$~mW.
}
\label{fig:dissociation_rempi_ions}
\end{figure}

We have measured the remaining ion fraction after a variable time of interaction $\Delta t_{d} = 0...10 \, \text{s}$ with the photodissociation laser of average power $P_d=9.8$~mW, for ions created in the $v=0$ (respectively  $v=1$) states by setting the REMPI laser to $\lambda = 302.47$~nm (respectively $\lambda = 295.61$~nm). In addition, a measurement was performed for $v=1$ at a higher power, $P_d=22.6$~mW. Results are shown in Fig.~\ref{fig:dissociation_rempi_ions}. We observe a fast decay of the ion population for ions created in the $v=1$ state, and a much slower decay for ions created in the $v=0$ state, as expected due to the 72 times smaller dissociation cross-section.

We performed a global fit of the data taken at power $P_d = 9.8$~mW using Eq.~\ref{eq:diss_model} with a single free parameter $\eta$. The vibrational populations measured in Ref.~\cite{OHalloran87} are incorporated in the fit model. These populations are respectively $n_0 = 0.90$, $n_1 = 0.10$ for $\lambda = 302.47$~nm, and $n_0 = 0.09$, $n_1 = 0.80$, $n_2 = 0.11$ for $\lambda = 295.61$~nm. Note that in the first case, a theoretical calculation of vibrational branching ratios yielded $n_0 = 0.96$, $n_1 = 0.04$~\cite{Dixit84}.

The fit is in good agreement with the data for $\eta \approx 0.16$. Using this value, the model correctly predicts the behavior for $P_d=22.6$~mW. This indicates that both the assumed vibrational populations~\cite{OHalloran87} and the calculated relative dissociation cross-sections~\cite{Busch72} provide a good description of the photodissociation dynamics. The value of $\eta$ is smaller than could be expected from the discussion of the previous section, which we attribute to imperfect alignment of the dissociation beam with the H$_2^+$ ions. Indeed, it may be explained by a mismatch of about 250~$\mu$m in the $y$ direction (see Fig.~\ref{fig:sketch} (c)) with respect to the trap axis, compatible with the uncertainty of our alignment procedure. This may be optimized further in the future.

Our results show that selective dissociation by a 213~nm laser is an effective method to distinguish the vibrational levels $v=0$ and $v=1$ in H$_2^+$, in a characteristic time on the order of 1~s with an average power $P_d=22.6$~mW.

Since the photodissociation cross-sections depend only weakly on the rotational state, we cannot  assess the rotational populations in the ion sample with this method. However, strong rotational selectivity of the REMPI process is expected due to the strong propensity rule of the ionization process~\cite{Tran13,Dixit85}. A previous study has observed qualitatively a high level of rotational selectivity (see Fig.~7 in Ref.~\cite{OHalloran87}).

\section{Conclusion}

By combining (3+1) REMPI from a pulsed H$_2$ molecular beam, and sympathetic cooling with laser-cooled Be$^+$ ions, we have demonstrated the production, trapping, and cooling of state-selected H$_2^+$ ions in the $v=0$ or $v=1$ vibrational levels. Differential pumping in the molecular beam apparatus allows a sufficiently low background pressure to be maintained in the trap's vacuum chamber. Deep-UV photodissociation of H$_2^+$ ions produced by REMPI has yielded results that are compatible with the high vibrational selectivity measured in previous studies, i.e. 90\% for the ground vibrational state $v=0$, and 80\% for $v=1$. Finally, we have shown that photodissociation at 213~nm
can be used to detect the excitation of H$_2^+$ ions from $v=0$ to $v=1$ with a rate of approximately 1/s. We have thus demonstrated all the necessary tools to prepare and detect H$_2^+$ ion samples in a high-resolution spectroscopy measurement.

Finally, the hyperfine state selectivity of the REMPI process~\cite{Germann16} is also an interesting topic for further investigations, with the perspective of producing hydrogen molecular ions in a single internal quantum state.

\section{Ackowlegments}
Thanks are due to A. Douillet for setting-up the 626~nm laser. M.S. acknowledges support by the Austrian Science Fund FWF within the DK-ALM: W1259-N27. This work has been supported by Region Ile-de-France in the framework of DIM SIRTEQ, by the ANR BESCOOL project, grant ANR-13-IS04-0002 of the French Agence Nationale de la Recherche, by the LABEX Cluster of Excellence FIRST-TF (ANR-10-LABX-48-01), within the Program "Investissements d'Avenir” operated by the French National Research Agency (ANR), and by the Initial Training Network COMIQ FP7-PEOPLE-2013-ITN grant 607491.

\end{document}